\documentclass[useAMS,usenatbib,usegraphicx]{mn2e}
\usepackage{natbib}
\usepackage{fixltx2e}
\usepackage{graphicx}
\usepackage{epstopdf} 
\usepackage{epsfig}
\usepackage{float}
\usepackage{amssymb}
\usepackage[fleqn]{amsmath}
\usepackage{multirow}
\usepackage{color}
\usepackage{caption}
\usepackage{subcaption} 
\usepackage{aecompl}

\newcommand{\affil}[1]{$^{\rm #1}$}
\newcounter{inst} 
\newcommand{\inst}[1]{\noindent%
   \refstepcounter{inst}\affil{\Alph{inst}\label{#1}}  
   }
\newcommand{\tabft}[1]{(#1)}
\newcounter{ft}
\newcommand{\ft}[1]{\noindent%
   \refstepcounter{ft}\tabft{\alph{ft}\label{#1}}}
\newcounter{todo}
\renewcommand\thetodo{\Alph{todo}}
\def\todo#1{\addtocounter{todo}{1}[[\thetodo: #1]]\strut\vadjust{%
\kern-\dp\strutbox{\vtop to \dp\strutbox{\baselineskip\dp\strutbox\vss\rlap{%
\hskip\hsize\ \rm{$\leftarrow$\thetodo}}\null}}}}
\def\LaTeX{L\kern-.36em\raise.3ex\hbox{a}\kern-.15em 
    T\kern-.1667em\lower.7ex\hbox{E}\kern-.125emX}

\newcommand{\arcdeg}{\ensuremath{^{\circ}}}
\title[Serendipitous discovery of a dying GRG
associated with NGC\,1534, using the MWA]{Serendipitous discovery of a dying Giant Radio Galaxy associated with NGC\,1534, using the Murchison Widefield Array}
\author[Hurley-Walker~et~al.]{Natasha Hurley-Walker\affil{\ref{ICRAR}},
Melanie Johnston-Hollitt\affil{\ref{VUW}},
Ron Ekers\affil{\ref{CASS},\ref{ICRAR}},
Richard Hunstead\affil{\ref{USyd}},\newauthor
Elaine M. Sadler\affil{\ref{USyd}},
Luke Hindson\affil{\ref{VUW}}, 
Paul Hancock\affil{\ref{ICRAR},\ref{CAASTRO}}, 
Gianni Bernardi\affil{\ref{SKASA},\ref{Rhodes},\ref{CfA}}, \newauthor
Judd D. Bowman\affil{\ref{ASU}}, 
Frank Briggs\affil{\ref{CAASTRO},\ref{ANU}}, 
Roger Cappallo\affil{\ref{Haystack}},
Brian Corey\affil{\ref{Haystack}}, \newauthor
Avinash A. Deshpande\affil{\ref{RRI}}, 
David Emrich\affil{\ref{ICRAR}}, 
Bryan M. Gaensler\affil{\ref{CAASTRO},\ref{USyd}},
Robert Goeke\affil{\ref{Haystack}}, \newauthor
Lincoln Greenhill\affil{\ref{CfA}},
Bryna J. Hazelton\affil{\ref{UDub}}, 
Jacqueline Hewitt\affil{\ref{MIT}}, 
David L. Kaplan\affil{\ref{UWM}}, \newauthor
Justin Kasper\affil{\ref{CfA}},
Eric Kratzenberg\affil{\ref{Haystack}}, 
Colin Lonsdale\affil{\ref{Haystack}}, 
Mervyn Lynch\affil{\ref{ICRAR}}, \newauthor
Daniel Mitchell\affil{\ref{CASS},\ref{CAASTRO}},
Russell McWhirter\affil{\ref{Haystack}}, 
Miguel Morales\affil{\ref{UDub}},
Edward Morgan\affil{\ref{MIT}}, \newauthor
Divya Oberoi\affil{\ref{NCRA}},
Andr\'{e} Offringa\affil{\ref{CAASTRO},\ref{ANU}},
Stephen Ord\affil{\ref{ICRAR},\ref{CAASTRO}},
Thiagaraj Prabu\affil{\ref{RRI}},
Alan Rogers\affil{\ref{Haystack}}, \newauthor 
Anish Roshi\affil{\ref{NRAO}},
Udaya Shankar\affil{\ref{RRI}},
K. Srivani\affil{\ref{RRI}},
Ravi Subrahmanyan\affil{\ref{RRI},\ref{CAASTRO}}, \newauthor
Steven Tingay\affil{\ref{ICRAR},\ref{CAASTRO}},
Mark Waterson\affil{\ref{ICRAR},\ref{ANU}},
Randall B. Wayth\affil{\ref{ICRAR},\ref{CAASTRO}},
Rachel Webster\affil{\ref{CAASTRO},\ref{UMelb}}, \newauthor
Alan Whitney\affil{\ref{Haystack}},
Andrew Williams\affil{\ref{ICRAR}},
Chris Williams\affil{\ref{MIT}}\\
{\small \affil{}\,Email: nhw@icrar.org}\\
{\small \inst{ICRAR}\,International Centre for Radio Astronomy Research, Curtin University, Bentley, WA 6102, Australia}\\
{\small \inst{VUW}\,School of Chemical \& Physical Sciences, Victoria University of Wellington, Wellington 6140, New Zealand}\\
{\small \inst{CASS}\,CSIRO Astronomy and Space Science, Marsfield, NSW 2122, Australia}\\
{\small \inst{USyd}\,Sydney Institute for Astronomy, School of Physics, The University of Sydney, NSW 2006, Australia}\\
{\small \inst{CAASTRO}\,ARC Centre of Excellence for All-sky Astrophysics (CAASTRO)}\\
{\small \inst{SKASA}\,Square Kilometre Array South Africa (SKA SA),3rd Floor, The Park, Park Road, Pinelands, 7405, South Africa}\\
{\small \inst{Rhodes}\,Department of Physics and Electronics, Rhodes University, PO Box 94, Grahamstown, 6140, South Africa}\\
{\small \inst{CfA}\,Harvard-Smithsonian Center for Astrophysics, 60 Garden Street, Cambridge, MA, 02138, USA}\\  
{\small \inst{ASU}\,School of Earth and Space Exploration, Arizona State University, Tempe, AZ 85287, USA}\\
{\small \inst{ANU}\,Research School of Astronomy and Astrophysics, Australian National University, Canberra, ACT 2611, Australia}\\
{\small \inst{Haystack}\,MIT Haystack Observatory, Westford, MA 01886, USA}\\
{\small \inst{RRI}\,Raman Research Institute, Bangalore 560080, India}\\
{\small \inst{UDub}\,Department of Physics, University of Washington, Seattle, WA 98195, USA}\\
{\small \inst{MIT}\,Kavli Institute for Astrophysics and Space Research, Massachusetts Institute of Technology, Cambridge, MA 02139, USA}\\
{\small \inst{UWM}\,Department of Physics, University of Wisconsin--Milwaukee, Milwaukee, WI 53201, USA}\\
{\small \inst{NCRA}\,National Centre for Radio Astrophysics, Tata Institute for Fundamental Research, Pune 411007, India}\\
{\small \inst{NRAO}\,National Radio Astronomy Observatory, Charlottesville and Greenbank, USA} \\
{\small \inst{UMelb}\,School of Physics, The University of Melbourne, Parkville, VIC 3010, Australia}\\
}
\begin{document}

\date{Accepted 0000 December 00. Received 0000 December 00; in original form 0000 October 00}

\pagerange{\pageref{firstpage}--\pageref{lastpage}} \pubyear{2014}

\maketitle

\label{firstpage}

\begin{abstract}
Recent observations with the Murchison Widefield Array at 185~MHz have serendipitously unveiled a heretofore unknown giant and relatively nearby ($z = 0.0178$) radio galaxy associated with NGC\,1534.
The diffuse emission presented here is the first indication that NGC\,1534 is one of a rare class of objects (along with NGC\,5128 and NGC\,612) in which a galaxy with a prominent dust lane hosts radio
emission on scales of $\sim$700\,kpc.
We present details of the radio emission along with a detailed comparison with other radio galaxies with disks. NGC1534 is the lowest surface brightness radio galaxy known with an estimated scaled 1.4-GHz surface brightness of just 0.2\,mJy\,arcmin$^{-2}$. The radio lobes have one of the steepest spectral indices yet observed: $\alpha=-2.1\pm0.1$, and the core to lobe luminosity ratio is $<0.1$\%.
We estimate the space density of this low brightness (dying) phase of radio galaxy evolution as $7\times10^{-7}$\,Mpc$^{-3}$ and argue that normal AGN cannot spend more than 6\% of their lifetime in this phase if they all go through the same cycle.

\end{abstract}


\section{Introduction}
\label{sec:introduction}

Giant Radio Galaxies (GRG) are characterised by larger linear size (usually $\geq700$\,kpc; see e.g. \citealt{2005AJ....130..896S}), and a radio spectrum that is much steeper than normal radio galaxies. These galaxies have a low surface brightness and without an active nucleus to provide fresh particle injection, they probably represent a short-lived, final stage of radio galaxy evolution \citep[e.g.][]{2011A+A...526A.148M}. The short timescale of this final phase of evolution makes these sources rare. The typically low surface brightness of GRGs means that only surveys with high brightness sensitivity are able to detect them.

Detection of GRGs at low ($\leq300$\,MHz) radio frequencies reveals details of the low-energy electron population present in the radio lobes which in turn provides information on the acceleration processes required to generate and maintain the emission.
Instruments such as the Low Frequency Array \citep[LOFAR;][]{2013A+A...556A...2V}, the Giant Metrewave Radio Telescope (GMRT) and the Murchison Widefield Array \citep[MWA;][]{2009IEEEP..97.1497L,2013PASA...30....7T} are thus important in the study of this class of object and have the potential to detect such sources in greater numbers.
In particular, LOFAR and MWA are expected to detect older ``dying'' GRGs whose radio lobes are comprised of old plasmas of low-energy electrons, whose steep spectra make them invisible at high radio frequencies. Here we present the serendipitous detection of one such previously-unknown GRG associated with the lenticular galaxy NGC\,1534.

The Murchison Widefield Array is a new low-frequency telescope operating at the Murchison Radio-astronomy Observatory in remote Western Australia. Operating between 80 and 300\,MHz, the MWA represents the first fully operational precursor instrument for the Square Kilometre Array (SKA). 

During Director's Discretionary Time (DDT) observations made in late August~2013, the bright radio source PKS\,$0408-65$ was observed at 185\,MHz as a phase calibrator. The full-width-half-maximum of the primary beam of the MWA is $\approx30^\circ$ so the $\approx900$\,deg$^2$ surrounding this source were also imaged during the routine calibration process. An unusual source of low surface-brightness extended emission was serendipitously detected, and further images were made from the available observations in order to determine its nature.

This paper is laid out as follows.
Section~\ref{sec:reduction} describes the observations, the calibration and imaging strategy used in data reduction, and the flux density calibration procedures, including correcting for the MWA primary beam.
Section~\ref{sec:results} describes the properties of the radio source using MWA and archival data.
Section~\ref{sec:discussion} discusses the impact of the detection of this source, and Section~\ref{sec:conclusion} concludes the paper with future observing prospects for radio galaxies of a similar nature.

We adopt a standard set of cosmological parameters throughout with $H_0 = 73$\,km\,s$^{-1}${Mpc}$^{-1}$, $\Omega_m = 0.27$ \& ${\Omega}_{\Lambda}=0.73$. At $z=0.017802\pm0.000017$ \citep{2002LEDA.........0P}, the luminosity distance of NCG\,1534 is 74.2\,Mpc, and 1~arcmin corresponds to 20.8\,kpc.

All position angles are measured from North through East (i.e. counter-clockwise). All equatorial co-ordinates are J2000. Figures use the ``cubehelix'' colour scheme \citep{2011BASI...39..289G}.

\section{Observations and data reduction}
\label{sec:reduction}

As detailed by \citet{2013PASA...30....7T}, the MWA consists of 128 32-dipole antenna ``tiles'' distributed over an area approximately 3~km in diameter. Each tile observes two instrumental polarisations, ``X'' (16~dipoles oriented East-West) and ``Y'' (16~dipoles oriented North-South).
The signals from the tiles are collected by 16~in-field receiver units, each of which services eight tiles.

During the observations made on 2013~Aug~31, 126 of the 128 tiles were functioning normally. As the potential GRG was not the primary focus of the observations, only 5.5\,minutes of data were recorded, in three separate 1.8-minute observations at UTC 19:37:28, 20:01:28 and 20:25:27. Data were collected at 0.5-second, 40-kHz resolution in a 30.72-MHz band centred at 184.9\,MHz (hereafter written as 185\,MHz). The compact antenna layout of the MWA gives it extremely high surface brightness sensitivity: $\sigma_T\approx20$\,mK (6\,mJy\,arcmin$^{-2}$) for these 5-minute observations. The large number of elements results in very good $u,v$-coverage, and thus a well-behaved synthesised beam and excellent image quality, even for such a short observing time.

As the observations were originally made for the purpose of calibration, they are easily calibrated by using a model of PKS\,$0408-65$ and running \texttt{bandpass}, an algorithm from the Common Astronomy Software Applications (\textsc{CASA}\footnote{http://casa.nrao.edu/}), to produce per-observation, per-antenna, per-polarisation, per-channel gains, which were then applied. The data are imaged using a standard \texttt{clean}, with a pixel resolution of 0\farcm75, an image size of $4000\times4000$\,pixels, and a threshold of $3\sigma\approx0.3$\,Jy\,beam$^{-1}$ (measured after an initial shallow clean). We phase to PKS\,$0408-65$, and ignore wide-field effects, which are not significant at the distance NGC\,1534 lies from this phase centre ($\approx3\arcdeg$). Most importantly, we use a ``robust'' or ``Briggs'' \citep{1995AAS...18711202B} weighting scheme, with a robustness parameter of $+1.0$, which offers a compromise between sensitivity to large-scale structure (a four-fold improvement in brightness sensitivity compared to uniform weighting) and a synthesised beam with reasonably well-suppressed sidelobes ($\lesssim10\%$ compared to $\lesssim15\%$ for a naturally-weighted beam and $\lesssim5\,\%$ for a uniformly-weighted beam).

Following \citet{2014arXiv1410.0790H}, we use a mosaicking strategy to combine our observations into a single image. As each observation is self-calibrated, and the ionospheric conditions were calm (i.e. no significant direction-dependent position distortions) there is no ionospheric blurring when the images are combined. The point spread function is the average of the synthesised beams of the three individual observations: $305\arcsec\times231\arcsec$, at a position angle of 139\fdg1.

Each snapshot was taken with a different delay setting on the MWA analogue beamformers; an analytic primary beam model was calculated for each setting and used to weight and correct the images during the mosaicking step. This results in a smoothly-varying flux scale across the map, as the primary beam model is not yet reliable at 185\,MHz, $40^\circ$ from zenith, but has no sharp or discontinuous features.

Following the method of \citet{2014MNRAS.445..330H} we account for this flux calibration error and determine a flux scale relative to known, strong sources. We select a sample of 14 suitable flux calibrators drawn from the Parkes compilation of radio sources \citep{1990PKS} (which is itself a compilation from several surveys, most notably \citet{1981MNRAS.194..693L} ). These sources were selected with the following criteria: be unresolved in
all MWA bands, have a flux density greater than 1\,Jy at 843\,MHz and be well-fitted by a simple power-law with no sign of curvature at low frequencies. For these sources, we fit the spectral profile between 408 and 5000\,MHz using the power-law relation $S\propto\nu^{\alpha}$. We then extrapolate this fitted power-law to 185\,MHz and scale the flux in our MWA map at the position of each flux calibrator accordingly. To account for the variation of the flux density scale across the image we perform cubic interpolation using the scale factors of our 14 flux calibrators to determine a position-dependent flux density scale in the region covered by the 14 flux calibrators ($\approx20$\,deg$^2$ around NGC\,1534). We verify the accuracy of the applied flux density scale by comparing the flux density estimates in our scaled 185\,MHz map to the predicted flux density at 185\,MHz for five test sources: PKS\,0400-613, PKS 0430-624, PKS\,0405-640, PMN\,J0340-6507 and PMN\,J0406-6050. We present the predicted and corrected flux densities for these sources in Table~\ref{tab:Flux_test}. Since the test flux densities themselves are extrapolations, and the observed error does not increase with angular distance from NGC\,1534, we expect the dominant source of error to be the assumption that both the flux calibration and test sources have power-law spectra below 400\,MHz. In this region of the sky there is currently no better option; we estimate the contribution of the residual flux scaling error to our flux density estimates to be $\sim10$\%.
%
%
%
\begin{table*}
\caption{Test sources to examine the accuracy of the flux re-scaling procedure described in Section~\ref{sec:reduction}.}
\begin{tabular}{lcccc}
\hline 
Source & \multicolumn{2}{c}{Flux Density}   & Residual & Angular distance \\
Name & Predicted  & Corrected   & difference & from NGC\,1534\\
 & (Jy) &  (Jy)  & (mJy) & (deg.) \\
\hline 
PKS\,0405-640 & 3.06 & 2.88 & $+18\pm6$  & 0.93 \\
PKS\,0400-613 & 1.66 & 1.35 &  $+31\pm19$  & 0.96 \\
PMN\,J0406-6050 & 1.86 & 1.50 & $+36\pm12$   & 2.06\\
PKS\,0430-624 & 3.64 & 3.76 & $-12\pm3$   & 2.75 \\
PMN\,J0340-6507 & 3.76 & 4.42 &  $-66\pm50$  & 4.86 \\
\hline 
\end{tabular}
\label{tab:Flux_test}
\end{table*}
\section{Results}
\label{sec:results}
\begin{figure*}
\begin{center}
\vbox{\includegraphics[width=17cm]{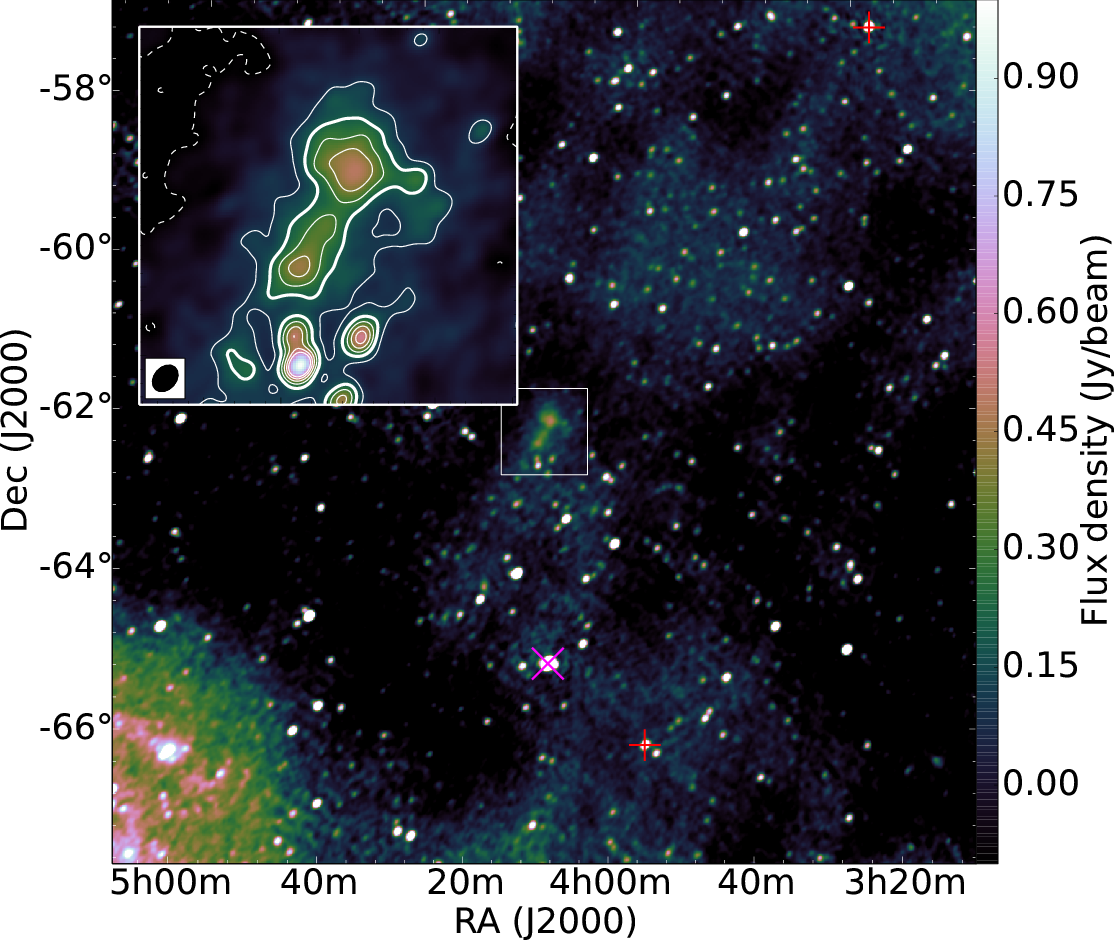}}
\caption{185\,MHz image of NCG\,1534, showing the surrounding 100\,deg$^2$ with a linear colourscale between $-0.1$ and $1.0$\,Jy\,beam$^{-1}$. The two sources marked with red ``$+$''s are two of the flux calibration sources described in Section~\ref{sec:reduction}. The phase calibrator, PKS\,$0408-65$ is marked with a magenta ``X''. The inset image shows a zoomed view of the radio lobes, with solid contours starting at $0.1$\,Jy\,beam$^{-1}$ and increasing in $\sigma=100$\,mJy\,beam$^{-1}$ increments to $+0.6$\,Jy\,beam$^{-1}$, a thicker contour at $2\sigma$, and a dashed contour at $-1\sigma$. The PSF is shown as a solid ellipse, of dimensions $305\arcsec\times231\arcsec$, at a position angle of 139\fdg1.
Henceforth, all images are shown with the same size and sky location as this inset image.\label{fig:NGC1534}}
\end{center}
\end{figure*}

Figure~\ref{fig:NGC1534} presents the robust $+1.0$ weighted MWA image of 10\% of the field, showing that the data have reasonably consistent noise characteristics. The Large Magellanic Cloud is to the lower-left and diffuse Galactic emission is evident in the upper right and lower middle of the image.
In the centre of the field we clearly see two diffuse patches of low-surface-brightness emission significantly above the Galactic emission in the image. The morphology
 is reminiscent of a double-lobed radio galaxy, but with wide lobes, and no obvious hot spots or core. The inset shows a zoom of the central source and two sources to the south; the eastern structure is revealed in higher-resolution radio data (e.g. the Sydney University Molonglo Sky Survey \citep[SUMSS;][]{1999AJ....117.1578B,2003MNRAS.342.1117M} at 843\,MHz) as two compact radio sources confused together by the large MWA synthesised beam, and the western structure is an FR-II radio galaxy aligned N-S.

It is not entirely straightforward to measure the noise properties of the map, given that it contains a large amount of diffuse structure on different scales.
We employ the technique of Hancock~et~al.~(in prep), in which background and noise images are determined using a spatial filter similar to a box-car filter. This is an efficient implementation ideal for large radio images compared to standard algorithms optimised for small optical images. The background image is calculated as the median of pixels within a rectangular region of size $20\times20$ synthesised beams ($\approx2$\,deg$^{2}$). The noise image is calculated from the inter-quartile range of the median subtracted pixels over the same region. Since radio images have a high spatial correlation, we calculate the median and inter-quartile range on a grid with a spacing of $4\times4$ pixels, and linearly interpolate to form a final image. We measure the background level as 5\,mJy\,beam$^{-1}$, and subtract it from our measured flux densities; the RMS is calculated as 100\,mJy\,beam$^{-1}$ in the vicinity of NGC\,1534.
(This agrees well with the more manual method of measuring the mean and RMS in nearby regions that have no obvious source peaks.)

We use the RMS measurement of $100$\,mJy\,beam$^{-1}$ as an estimate of the noise in the image, and measure the flux density of the radio galaxy lobes by integrating down to multiples of 2 and 3 times this value. 
First, we measure the total flux density of the source by integrating down to 2\,$\sigma$. We then measure the relative power of the northern and southern lobes by integrating each lobe down only to 3\,$\sigma$. We multiply this ratio by the 2-$\sigma$ integrated flux density to obtain the flux densities of the two lobes. We estimate the total 185\,MHz flux density of the radio lobes to be $4.8\pm0.5$\,Jy and the flux densities to be $3.0\pm0.3$ and $1.8\pm0.2$\,Jy for the southern and northern lobes, respectively (Table~\ref{tab:flux_density_measurements}).

Measured from the 2-$\sigma$ contours to the NW and SE, the angular extent of the lobes is 33~arcmin, or 700\,kpc in our cosmology. They lie at an angle of $334^\circ$ from North through East.

\begin{table*}
\centering
\caption{Integrated flux densities for the MWA, MRCR, SUMSS and CHIPASS data, after subtraction
of contaminating point sources and masking of negative pixels. Errors on the MWA flux densities may be taken as 10\,\%, due to the dominant calibration error from the flux calibration process described in Section~\ref{sec:reduction}. The SUMSS error bars are calculated as the RMS of the local region in the smoothed map. The MRCR data point with a ``*'' is a simple doubling of the upper limit given the morphology shown in Figure~\ref{fig:MRC}.\label{tab:flux_density_measurements}}
\begin{tabular}{|c|c|c|c|c|}
\hline 
\multirow{2}{*}{Instrument} & Frequency /  & \multicolumn{3}{c|}{Flux density / mJy}\tabularnewline
 & MHz & NW & SE & Total\tabularnewline
\hline 
MWA & 185 & $3000\pm300$ & $1800\pm180$ & $4800\pm480$ \tabularnewline
MRCR & 408 & $>260$ & $>150$ & $>410, \sim820*$\tabularnewline
SUMSS & 843 & $130\pm20$ & $80\pm20$ & $210\pm20$\tabularnewline
CHIPASS & 1400 & $<45$ & $<27$ & $<72$\tabularnewline
\hline 
\end{tabular}
\end{table*}
\subsection{Host galaxy}
\begin{figure}
\begin{center}
\vbox{\includegraphics[width=8.75cm]{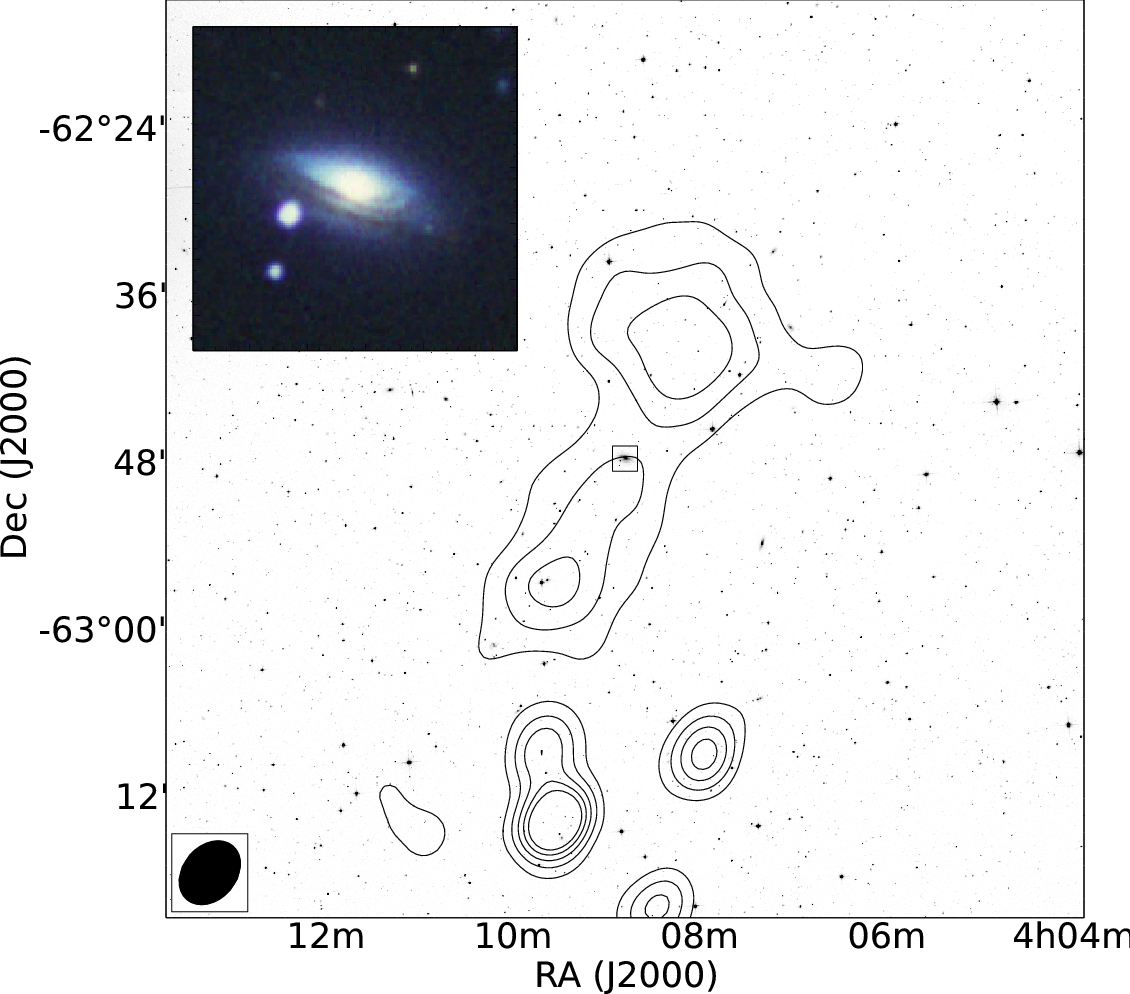}}
\caption{MWA 185-MHz contours starting at $2\sigma$ and proceeding in $\sigma=100$\,mJy increments, overlaid on the Digitized Sky Survey (DSS) Red image. NGC\,1534 can be seen at the top of the southern lobe. The inset shows a false color image of NGC\,1534, marked on the main image with a square; red represents the UK Schmidt infra-red image, green the the European Southern Observatory Red image, and blue the UK Schmidt Blue image. A dust lane is clearly visible.}\label{fig:NGC1534_DSS}
\end{center}
\end{figure}
\begin{table*}
\caption{Properties of the host galaxy NGC\,1534.\label{tab:NGC1534_optical}}
\begin{tabular}{|c|c|c|}
\hline 
Property & Value & Reference\tabularnewline
\hline 
Position (J2000) & $04^{\mathrm{h}}08^{\mathrm{m}}46.066^{\mathrm{s}}$ $-62^{\circ}47\arcmin51\farcs21$ & \citet{2006AJ....131.1163S}\tabularnewline
Redshift & $0.017802\pm0.000017$ & \citet{2002LEDA.........0P}\tabularnewline
Morphological type & SA0/a?(rs) & \citet{1991rc3..book.....D}  \tabularnewline
IRAS 60$\mu$m flux density & $0.332\pm0.033$\,Jy & \citet{1984ApJ...278L...1N} \tabularnewline
IRAS 100$\mu$m flux density & $1.885\pm0.188$\,Jy & \citet{1984ApJ...278L...1N} \tabularnewline
Position angle of dust lane (N through E) & $250^\circ$ & \citet{2006AJ....131.1163S} \tabularnewline
\textsc{Hi} mass & $<1.0\times10^{10}$\,M$_\odot$ & \cite{2001MNRAS.322..486B} \tabularnewline
$M_K$ & $-25.0$ & \citet{2006AJ....131.1163S} \tabularnewline
\hline 
\end{tabular}
\end{table*}
Figure~\ref{fig:NGC1534_DSS} shows the MWA radio contours overlaid on the Digitized Sky Survey (DSS) red field. At the centre of these two diffuse patches we see a relatively bright 13.7\,mag galaxy at a redshift of 0.0178, NGC\,1534: see Table~\ref{tab:NGC1534_optical}. The galaxy has an obvious disk component with a dust lane as well as a dominant nuclear bulge. \citet{1991rc3..book.....D} classify it as SA0/a?(rs) edge-on: a lenticular, with the possibility of being a spiral. Spiral structure is somewhat visible in the optical image. \citet{2007ApJ...655..790C} associate NGC\,1534 with the galaxy groups LDC\,292 and HDC\,269, which have 17 and 3 members, respectively.

The position angle of the lobes is well-aligned with the minor axis of the host galaxy dust lane (Table~\ref{tab:NGC1534_optical}), with a difference of only $6^\circ$; this is consistent with other GRGs which exhibit such minor axis alignment \citep{2009ApJ...695..156S} and supports NGC\,1534 as the host galaxy. We note that there is no radio source coincident with NGC\,1534 itself, at any frequency.

\subsection{Other radio images}

\begin{figure*}
\centering
    \begin{subfigure}[b]{0.48\textwidth}
                \includegraphics[width=\textwidth]{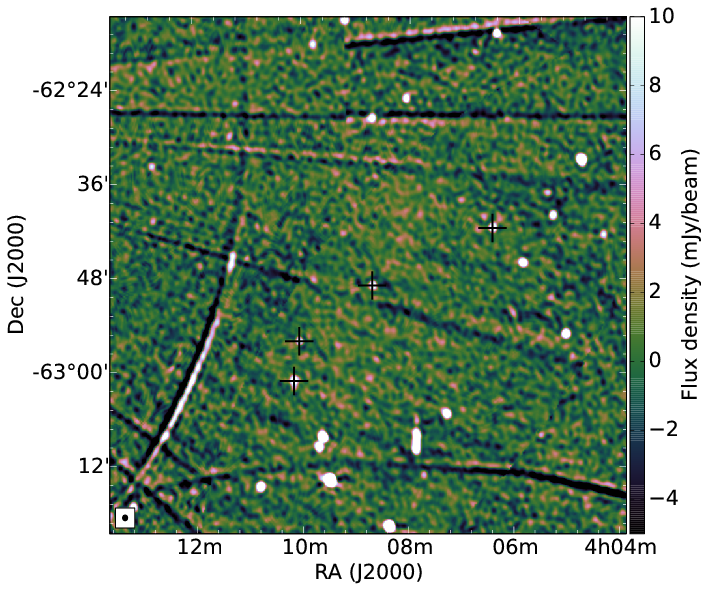}
    \end{subfigure}
    \begin{subfigure}[b]{0.48\textwidth}
                \includegraphics[width=\textwidth]{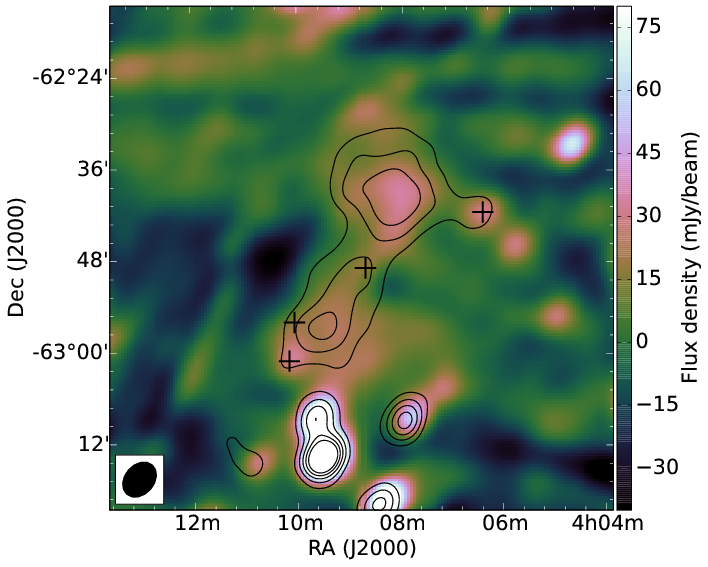}
    \end{subfigure}
\caption{SUMSS 843\,MHz image of the NCG\,1534 region with the four SUMSS catalogue sources indicated with black crosses. The flux densities of these sources are given in Table~\ref{tab:contaminating_sources}. The left panel shows the original SUMSS image, and the right panel shows the image after convolution to match the MWA resolution. The arc-like artifacts in the SUMSS image are grating responses from the bright sources PKS\,0408$-$65 and PKS\,0420$-$62, and
   the linear features are radial artifacts from PKS\,0420$-$62. The synthesised beams of SUMSS ($48\arcsec\times43\arcsec$; position angle $0^\circ$) and the MWA ($305\arcsec\times231\arcsec$; position angle 139\fdg1) are shown as filled ellipses in the lower-left corner of the left and right panels, respectively. On the right panel, MWA 185-MHz contours start at $2\sigma$ and proceed in $\sigma=100$\,mJy increments.
\label{fig:SUMSS}}
\end{figure*}

The Sydney University Molonglo Sky Survey \citep[SUMSS;][]{1999AJ....117.1578B,2003MNRAS.342.1117M} at 843\,MHz is the highest-resolution low-frequency radio catalogue covering this region. The SUMSS image shows four contaminating point sources within the envelope of the diffuse radio emission, whose positions and fluxes are given in Table~\ref{tab:contaminating_sources}, and whose positions are indicated by ``$+$''s in the left panel of Figure~\ref{fig:SUMSS}. 
The contaminating point sources are unresolved by all other instruments discussed in this paper. An average spectral index, $\alpha=-0.7$ was used to estimate the flux densities at other frequencies, and these were subtracted from any flux density measurement of the lobes which overlapped with their positions.

To highlight the small amount of structure on scales~$>0\fdg5$ detected by SUMSS, we convolve the image to match the MWA synthesised beam. The right panel of Figure~\ref{fig:SUMSS} shows the resulting image with MWA 185-MHz contours overlaid: the radio galaxy lobes are just visible. We calculate the integrated flux density as 209\,mJy by summing the flux under the integration contours derived from the MWA data, and subtracting the contaminating sources as described above. The RMS noise on the smoothed SUMSS image is $\approx20$\,mJy in this area so we ascribe a $10$\% error to this flux density measurement.

The Molonglo Reference Catalogue \citep[MRC;][]{1981MNRAS.194..693L} at 408\,MHz also covers this region but only reaches a completeness flux density of 0.95\,Jy. However, the survey itself goes far deeper, with an RMS noise level of $\sim$30--50 mJy, and a revised version known as MRCR (described in \citealt{2007MNRAS.381..341B}) has been created in image form (D.F. Crawford, private communication). The MRCR image in the vicinity of MWA J0408$-$6247 is unfortunately contaminated by the north-south sidelobes of PKS\,$0408-65$ ($S_{\mathrm{408\,MHz}} = 51$\,Jy), which is $3^{\circ}$ due south, and attempts at deconvolution have left a negative residue.
To place a lower limit of 410\,mJy, on the flux density detected in the image, we masked all pixels of negative values, integrated under the same MWA contours, and subtracted the estimated contaminating source flux densities. As the side lobe cuts the source in half we can assume to first order that the total flux density is approximately 820\,mJy. 

We note that the shortest baseline of the Molonglo Cross is 15\,m, so the MRCR should be sensitive to scales up to 2\fdg8, and SUMSS to scales up to 1\fdg4, which is sufficient to measure the flux density of the radio galaxy lobes, given the necessary sensitivity and the absence of image processing artefacts.

By reprocessing archival data from the HI Parkes All-Sky Survey (HIPASS), \citet{2014PASA...31....7C} present a 1.4\,GHz continuum map of the sky south of Declination $+25^\circ$ (CHIPASS). At the location of the radio lobes, the continuum 1.4\,GHz (CHIPASS) image shows no obvious maximum, and the peak brightness temperature is 3.6\,K, similar to the background temperature of the whole region. Given that we know the location of the expected emission from the radio lobes, we use the $2\sigma$ sensitivity of the image to estimate an upper limit on the flux density of the radio lobes at 1.4\,GHz. The $1\sigma$ RMS is 18\,mJy\,beam$^{-1}$ and the radio lobes occupy two 14\farcm4-FWHM Parkes beams, so an upper ($2\sigma$) flux density limit at 1.4\,GHz is 72\,mJy.

At 4.85\,GHz, the Parkes-MIT-NRAO (PMN) survey \citep{1993AJ....106.1095C} shows no detection of either a source at NGC\,1534's position or of the diffuse radio lobes. Following a similar deduction as with CHIPASS, the $1\sigma$ RMS is 8\,mJy\,beam$^{-1}$ and the radio lobes occupy twelve 5~arcmin-FWHM Parkes beams, so an upper ($2\sigma$) flux density limit at 4.85\,GHz is 200\,mJy.

\begin{figure}
\begin{center}
\vbox{\includegraphics[width=8.75cm]{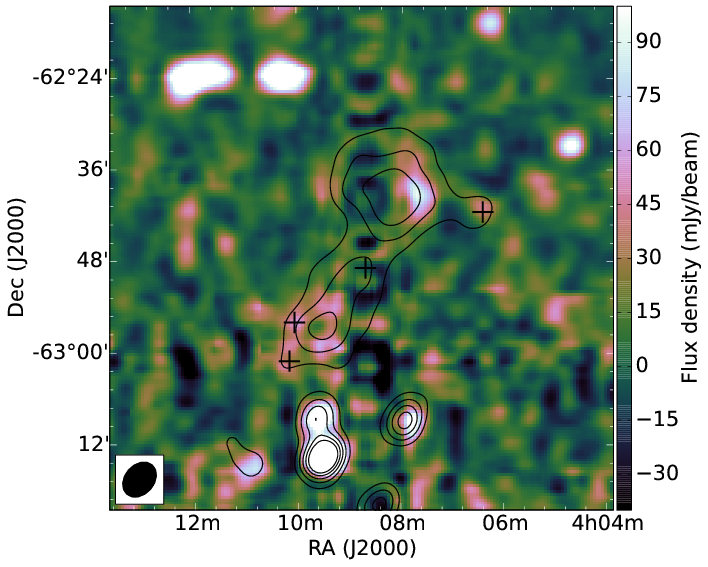}}
\caption{
 408 MHz image of the NGC\,1534 region from MRCR. MWA 185-MHz contours start at 2$\sigma$ and
   proceed in $\sigma = 100$\,mJy increments; the MWA synthesised beam is shown as a filled ellipse in the lower-left corner. The MRCR beam is $3\farcm2 \times 2\farcm6$ (position angle $0^{\circ}$).  Unfortunately, the N-S sidelobes of PKS\,0408$-$65 bisect the lobes of NGC\,1534 and leave a negative residue. The two bright ``sources'' to the NE are E-W sidelobes of PKS\,0420$-$62.
%
%
The MWA synthesised beam is shown as a filled ellipse in the lower-left corner.\label{fig:MRC}}
\end{center}
\end{figure}
%
\begin{table*}
\centering
\caption{Sources detected within the radio lobes as measured by the SUMSS radio survey. The two columns with ``*''s show extrapolated flux densities at the MRCR and MWA observing frequencies using a spectral index $\alpha$ of $-0.7$. One source is not subtracted from the MRCR data, as it lies in the region of the poorly-subtracted grating lobe, so we do not list its predicted flux density.\label{tab:contaminating_sources}}
\begin{tabular}{|c|c|c|c|c|}
\hline
  \multicolumn{1}{|c|}{RAJ2000} &
  \multicolumn{1}{c|}{DEJ2000} &
  \multicolumn{1}{c|}{$S_{\mathrm{843MHz}}$ (mJy) }&
  \multicolumn{1}{c|}{$S^{*}_{\mathrm{408MHz}}$ (mJy) }&
  \multicolumn{1}{c|}{$S^{*}_{\mathrm{185MHz}}$ (mJy) } \\
\hline
  04 06 27.54 & $-62$ 41 44.50 & $21.7\pm1.1$ & $36.1\pm1.8$ & $64.0\pm3.2$ \\
  04 08 41.52 & $-62$ 49 09.20 & $17.5\pm1.1$ & -- & $51.6\pm3.2$ \\
  04 10 03.30 & $-62$ 56 15.20 & $12.4\pm1.1$ & $20.6\pm1.8$ & $36.5\pm3.2$ \\
  04 10 09.43 & $-63$ 01 19.30 & $27.8\pm2.6$ & $46.2\pm4.3$ & $81.9\pm7.7$ \\
\hline\end{tabular}
\end{table*}
In summary, the MWA is the only instrument with the resolution and surface brightness sensitivity capable of making a reliable detection of the radio lobes, and their presence in the Molonglo data provides a useful flux density estimate at a higher frequency. 

\subsection{Radio structure and spectrum}

\begin{figure}
\begin{center}
\includegraphics[width=8.75cm]{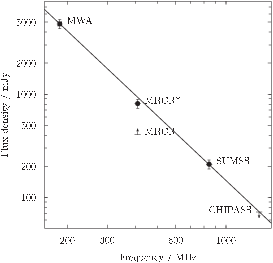}
\caption{Plot of the measured values of the flux density of the radio lobes of NGC\,1534 at 185, 408, 843 and 1400\,MHz, from MWA, MRCR, SUMSS and CHIPASS, respectively, as shown in Table~\ref{tab:flux_density_measurements}. The line indicates a least-squares unweighted fit made to the MWA and SUMSS measurements and CHIPASS upper limit, with a spectral index of $\alpha=-2.1\pm0.1$.}\label{fig:spectrum}
\end{center}
\end{figure}

NGC\,1534 has extreme radio properties. In the MWA images the lobes are very diffuse with no sign of any jets or hot spots. Fitting a power law spectrum ($S\propto\nu^\alpha$) to the MWA and SUMSS measurements, and the CHIPASS upper limit (see Figure~\ref{fig:spectrum}), we derive a spectral index $\alpha=-2.1\pm0.1$ for the lobes. This is much steeper than typical radio galaxies with active central jets. We note that the CHIPASS flux density could be lower, which would lead to a spectrum with a break.

No nuclear source is detected by the MWA and the peak brightness at NGC\,1534's position is 270\,mJy\,beam$^{-1}$. We look to ancillary data to place limits on the core flux density.

No source is visible in the Australia Telescope 20\,GHz survey \citep{2010MNRAS.402.2403M}, and the RMS noise level at its position is 5.7\,mJy\,beam$^{-1}$ (Hancock~2014, priv. comm.), so we give a 2$\sigma$ upper limit of 11\,mJy. Similarly, given an RMS noise level of 7.8\,mJy\,beam$^{-1}$ in the 4.85\,GHz PMN survey, a 2$\sigma$ upper limit of 15.6\,mJy\,beam$^{-1}$ can be placed.
The peak brightness in the unsmoothed SUMSS data is 2.07\,mJy\,beam$^{-1}$. Initial exploratory observations made with the Australia Telescope Compact Array at 1.4GHz result in a 2$\sigma$ upper limit of 0.2\,mJy\,beam$^{-1}$. Further observations of the core will be presented in Johnston-Hollitt et al. (in prep).

Using the 1.4\,GHz measurement as the strongest limit, and assuming a typical flat-spectrum core, we scale to 843\,MHz in order to compare with the results of \citet{2005AJ....130..896S}. We find that the core is responsible for $<0.1$\% of the total flux density at 843\,MHz. This is a factor of two fainter than the faintest cores compiled by \citeauthor{2005AJ....130..896S}, implying very little emission from an active accretion disk.
\begin{table*}
\caption{Properties of known disk galaxies hosting large (linear size $>100$\,kpc) radio lobes. $L_{325\mathrm{MHz}}$ and $\alpha$ refer to the 325-MHz luminosity and spectral index of the radio lobes.\label{tab:gals_with_lobes}}
\begin{tabular}{|c|c|c|c|c|c|}
\hline 
Name & $z$ & Linear size (kpc) & Total $L_{325\mathrm{MHz}}$ (W\,Hz$^{-1}$) & $\alpha$ & $M_K$ \tabularnewline
\hline
2MASX~J\,23453268-0449256 & 0.07860 \tabft{\ref{2007M}} & 1580 \tabft{\ref{2014B}} & $5.2\times10^{25}$ \tabft{\ref{2014B}} & $-2.0$ \tabft{\ref{2014B}} & $-26.0$ \tabft{\ref{2006S}}\tabularnewline
NGC\,612 (PKS\,0131-36) & 0.030 \tabft{\ref{2002P}} & 500 \tabft{\ref{2008E}} & $3.2\times10^{25}$ \tabft{\ref{1985W}} & $-0.51$ \tabft{\ref{1985W}} & $-25.9$ \tabft{\ref{2006S}} \tabularnewline 

SDSS~J\,140948.85-030232.5 (Speca) & 0.14 \tabft{\ref{2009A}} & 1300 \tabft{\ref{2011H}} & $2.3\times10^{25}$\tabft{\ref{2011H}}& $-0.8$\tabft{\ref{2011H}} & $-25.6$ \tabft{\ref{2006S}} \tabularnewline
NGC\,1316 (Fornax~A) & 0.00591 \tabft{\ref{2002P}} & 389 \tabft{\ref{2014M}} & $2.2\times10^{25}$ \tabft{\ref{2014M}} & $-0.77$ \tabft{\ref{2014M}} & $-26.3$ \tabft{\ref{2006S}} \tabularnewline
NGC\,5128 (Centaurus~A) & 0.0018 \tabft{\ref{2011L}} & 500 \tabft{\ref{1965C}} & $5.5\times10^{24}$ \tabft{\ref{1965C}} & $-0.64$ \tabft{\ref{2013M}} & $-25.4$ \tabft{\ref{2006S}} \tabularnewline
6dFGS~gJ031552.1-190644 (0313-192) & 0.067 \tabft{\ref{1995O}} & 200 \tabft{\ref{1998L}} & $2.5\times10^{24}$ \tabft{\ref{1998L}} & unknown & $-24.7$ \tabft{\ref{2006S}} \tabularnewline
NGC\,1534 & 0.018 \tabft{\ref{2002P}} & 700 \tabft{\ref{2014H}} & $9.7\times10^{23}$ \tabft{\ref{2014H}} & $-2.1$ \tabft{\ref{2014H}} & $-25.0$ \tabft{\ref{2006S}} \tabularnewline
\hline 
\end{tabular}
\footnotesize{References:
\ft{2007M}~\cite{2007MNRAS.375..931M};
\ft{2014B}~\cite{2014ApJ...788..174B};
\ft{2006S}~\cite{2006AJ....131.1163S};
\ft{2002P}~\cite{2002LEDA.........0P};
\ft{2008E}~\cite{2008MNRAS.387..197E};
\ft{1985W}~\cite{1985MNRAS.216..173W};
\ft{2009A}~\cite{2009yCat.2294....0A}}
\ft{2011H}~\cite{2011MNRAS.417L..36H};
\ft{2014M}~\cite{2014arXiv1411.1487M};
\ft{2011L}~\cite{2011MNRAS.416.2840L};
\ft{1965C}~\cite{1965AuJPh..18..589C};
\ft{2013M}~\cite{2013MNRAS.436.1286M};
\ft{1995O}~\cite{1995AJ....109...14O};
\ft{1998L}~\cite{1998ApJ...495..227L};
\ft{2014H}~This~paper.
\end{table*}
\section{Discussion}\label{sec:discussion}
\subsection{Radio galaxies with disks}
Highly extended radio sources hosted by galaxies with disks are very rare, with only six other clear examples known:
\begin{itemize}
\item \textbf{J\,2345-0449}, recently discovered to be the largest radio galaxy associated with a spiral \citep{2014ApJ...788..174B};
\item \textbf{Speca} \citep[SDSS~J\,140948.85-030232.5;][]{2011MNRAS.417L..36H}, which has an inner steep-spectrum double and an outer relic lobe, part of which is re-accelerated and flatter-spectrum;
\item \textbf{PKS\,0131-36} \citep[NGC\,612;][]{2008MNRAS.387..197E}, which posseses a huge disk of cool \textsc{Hi} gas distributed along the optical disk and dust lane;
\item \textbf{Fornax~A} \citep[NGC\,1316;][]{1983A+A...127..361E}, whose dust lane is harder to see due to its face-on orientation;
\item \textbf{Centaurus~A} \citep[NGC\,5128;][]{1949Natur.164..101B}, which was recently imaged in detail by \citet{2011ApJ...740...17F};
\item \textbf{0313-192} (PMN\,J0315-906) \citep[6dFGS~gJ031552.1-190644;][]{2001ApJ...552..120L}, a 200\,kpc double-lobed radio galaxy with a disk.
\end{itemize}
In Table~\ref{tab:gals_with_lobes} we give the physical properties of these objects, and do not include the numerous examples of radio galaxies hosted by elliptical galaxies with small-scale circum-nuclear disks, or disk galaxies with smaller-scale emission such as B2\,0722+30, a \textsc{Hi}-rich spiral galaxy with $\approx14$\,kpc radio jets aligned in a manner suggesting tidal interactions with other nearby galaxies \citep{2009MNRAS.396.1522E}.

It has always been notable that the powerful double-lobed radio galaxies are identified with giant elliptical galaxies and are almost never found associated with spiral galaxies \citep{2010A+A...518A..10V}. However, those radio galaxy hosts with disks or dust lanes are particularly interesting since they make it possible to trace the dynamics of the gaseous material which is assumed to be the source of fuel in these systems. For instance, in the case of NGC\,612, \citet{2008MNRAS.387..197E} discovered an enormous disk of cool gas with $M_\textsc{Hi}=1.8\times10^9M_\odot$ distributed along the optical disk and dust lane. NGC1534 has a prominent dust lane, and hence must have a gaseous disk, but it is not yet detected in \textsc{Hi}, with a mass limit from the HIPASS survey of $<1.0\times10^{10}$\,M$_\odot$.

Using the infrared flux densities (Table~\ref{tab:NGC1534_optical}) measured by Infrared Astronomical Satellite (IRAS), and the infrared luminosity-dust temperature relation (Equation~3) of \citet{1989ApJS...70..699Y} adapted to IRAS 60- and 100-$\mu$m flux densities in Equation~7 of \citet{1991ApJS...75..751R}, we derive a characteristic dust 
temperature of $24\pm2$\,K. This is quite cool: generally $T_\mathrm{dust} > 30$\,K in galaxies with moderate on-going star formation. The estimated dust mass is $1.8\times10^7$\,M$_\odot$, which is similar to that in other early-type galaxies with prominent dust lanes. Given the typical gas/dust ratio for early-type galaxies with dust lanes, we would expect to see of order $10^{10}$\,M$_\odot$ of \textsc{HI} associated with this cool dust, which suggests that the current upper limit must be close to the actual value, and this galaxy ought to be detectable in \textsc{HI} with only a modest amount of additional observing time. Furthermore, the \textsc{HI} morphology and kinematics might give some additional hints about the past history of this galaxy. 

\begin{figure}
\centering
\includegraphics[width=8.75cm]{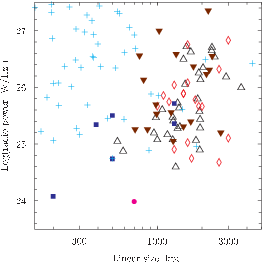}
\caption{The 325-MHz radio power against linear size $P-D$ diagram for GRGs in the literature: FR-I and FR-II objects compiled by \citet{1983MNRAS.204..151L} with $z<0.6$, scaled by a spectral index of $-0.8$ following \citet{2001A+A...374..861S} (plusses), GRGs discovered before 1998 (diamonds), GRGs discovered by \citet{2001A+A...374..861S} (empty triangles), GRGs measured by \citet{2005AJ....130..896S} (filled triangles), the radio lobes of NGC\,1534 as described in this paper (filled circle) and the six other radio galaxies hosted by galaxies with disks whose properties are given in Table~\ref{tab:gals_with_lobes} (filled squares).\label{fig:P-D}}
\end{figure}

\citet{2001A+A...374..861S} have compiled a sample of Giant Radio Galaxies from the Westerbork Northern Sky Survey (WENSS; \citealt{1997A+AS..124..259R}) and a survey of the literature. Rescaling our 185\,MHz flux density to 325\,MHz, assuming $\alpha=-2.1$, we calculate the NGC\,1534 radio jet luminosity at 325\,MHz as $9.7\times10^{23}$\,W\,Hz$^{-1}$. We replot Figure~3 of \citeauthor{2001A+A...374..861S} with our cosmology in Figure~\ref{fig:P-D}, and see that NGC\,1534 is inaccessible to a survey with the sensitivity of WENSS; \citeauthor{2001A+A...374..861S} note that low-redshift objects of linear size less than 1\,Mpc cannot be detected due to the low angular size limit of WENSS, while at higher redshifts they become faint and lie below the sensitivity threshold ($\approx10^{25}$\,W\,Hz$^{-1}$ for $z=0.1$). The lobes of NGC\,1534 are considerably fainter and larger than the typical FR-I and II radio galaxies compiled by \citet{1983MNRAS.204..151L}, which are powerful sources with active nuclei. We also include in the figure the eighteen GRGs discovered by \citet{2005AJ....130..896S} using SUMSS, by scaling their 843\,MHz luminosities to 325\,MHz with a spectral index of $-0.85$.

The radio galaxies with disks (Table~\ref{tab:gals_with_lobes}) are plotted in Figure~\ref{fig:P-D} for comparison. They cover the same range of linear sizes but have somewhat lower radio luminosities, excepting the recently-discovered J\,2345-0449. Their generally lower luminosities are probably a selection effect because the dust lanes and disks in early type galaxies may not be so easily identified in the more distant and more powerful sources.

The giant radio galaxy J\,0034.0-6639, observed in the Australia Telescope Low Brightness Survey \citep[ATLBS;][]{2012ApJS..199...27S} had the lowest surface brightness lobes known: 1\,mJy\,arcmin$^{-2}$ at 1.4\,GHz. J\,0034.0-6639 is 17~arcmin in extent, which is half the angular size of NGC\,1534, but at a redshift of 0.11 it is six times more distant and hence three times our linear size at 2\,Mpc. It has quite similar radio morphology to NGC\,1534 and is aligned with the minor axis of an E galaxy. Scaled to 1.4\,GHz assuming $\alpha=-2.1$, we calculate that the radio lobes of NGC\,1534 have a surface brightness of less than 0.1\,mJy\,arcmin$^{-2}$, making it the lowest-surface-brightness giant radio galaxy detected, by an order of magnitude.

\subsection{A relic giant radio galaxy?}

The lobes hosted by NGC\,1534 are certainly in the regime which makes them dim enough to be a true relic, but are physically smaller than the lobes of the GRGs compiled by \citet{2001A+A...374..861S}. As `relic' radio galaxies age, if the electrons are not re-accelerated, the jets become large and dimmer, and the spectral index steepens. The spectral index of the lobes of NGC\,1534, $\alpha=-2.1$, is comparable to the steepest spectra seen in other dying radio sources \citep[e.g.][]{2011A+A...526A.148M}. The lack of a central source implies that AGN activity has halted, sufficiently long ago for the lobes to age. Is there too little gas available as fuel, or is none being transported into the central black hole?

\citet{2011A+A...526A.148M} have shown that relic sources selected by their steep spectra are much more likely to be found in dense environments. This is important because when an AGN ceases to be active, it is expected that the lobes will expand and cool, but this is only the case in under-dense environments. For such sources found in environments of medium to high density, such as galaxy clusters, the intra-cluster medium is expected to prevent much expansion, effectively maintaining the lobe. Thus, in the early stages after AGN activity ceases, the observational properties of the lobes of `dead' radio galaxies in dense environments may vary little from those that are still active, whereas in under-dense regimes the lobes may freely expand giving rise to significantly dimmer objects with very steep spectral indices over similar timescales. Precise comparisons, however, require detailed knowledge of the environment (including the properties of ambient and lobe magnetic fields) and an understanding of the age of the emission. Although NGC\,1534 is associated with a widely dispersed galaxy group, it is unlikely to lie within a medium capable of preventing the lobe expansion; this is, again, consistent with it being a true `relic' galaxy. 

\subsection{Other possible origins?}

The physical size and proximity of the radio source to the nearest catalogued galaxy cluster, A\,3229 \citep{1989ApJS...70....1A} suggest that the emission might not be a GRG, but rather a radio relic on the outskirts of the cluster. However, investigations into A\,3229 suggest that it is not, in fact, a galaxy cluster. \citet{1989ApJS...70....1A} do not assign a redshift in the original identification; \citet{1991ApJ...383..467M} assign it a redshift of 0.017 based on a cross-match with the object DC\,0410-62 identified by \citet{1980ApJS...42..565D}, while it is reported in the NASA Extragalactic Database as having redshift 0.091923, based on a cross-match with the initial 6dF data release \citep{2004MNRAS.355..747J}. Examining the redshifts of the nearby galaxies as given by the final version of the 6dF galaxy survey \citep{2009MNRAS.399..683J} shows that there is no local overdensity of galaxies at $z\approx0.09$, and that the nearest true galaxy cluster is A\,3266, part of the Horologium-Reticulum Supercluster (HRS) \citep{2005AJ....130..957F}. The HRS is one of the most massive galaxy concentrations in the local Universe and is expected to host many dynamical events which might generate radio relics. However, for the observed lobes to be relic emission associated with the nearest cluster, they would need to lie 12\,Mpc away, which puts the emission beyond the limit of where simulations predict large-scale structure formation shocks can occur \citep{2001ApJ...562..233M}.

As previously noted, \citet{2007ApJ...655..790C} identify NGC\,1534 as belonging to the galaxy group HDC\,269 at $z=0.017$. There is no sign of X-ray emission in the R{\"O}ntgenSATellit \citep[ROSAT]{1984PhST....7..209T} All-Sky Survey data at its position, which, given the nearby redshift of the HRS, confirms this as a group, rather than a cluster.
We conclude that A\,3229 itself was misidentified as a cluster due to the chance alignment of HDC\,269 and the outskirts of the Horologium-Reticulum Supercluster, and that the lobes seen around NGC\,1534 cannot be relic emission from a cluster of galaxies.

Alternatively, the emission could be associated with a more distant radio galaxy.
The general alignment between the radio lobes and NGC\,1534 is quite reasonable (the position angles are nearly perpendicular, differing by $84^\circ$), although there are no inner jets or nuclear source to confirm the identification. Assuming the cosmology given in Section~\ref{sec:introduction}, the linear extent of the radio lobes is $\approx700$\,kpc. If the identification with NGC\,1534 is incorrect it must be associated with a more distant galaxy and the linear size would be much more extreme, and its radio power would be correspondingly larger.

\subsection{The population of low surface brightness radio galaxies}
At the time of the discovery of the lobes surrounding NGC\,1534, around 4,000\,deg$^2$ of sky had been imaged by the MWA with similar sensitivity to the observations described here, and examined to the same level of attention. This does not include the commissioning observations performed by \citet{2014arXiv1410.0790H} as they lacked the surface brightness-sensitivity of the full array. In this entire sky area, only one such object has been detected, although a concerted search of the full sky area accessible to the MWA has not yet begun. We note that the GaLactic and Extragalactic All-sky MWA survey (GLEAM; Wayth~et~al. in prep) will integrate for a similar amount of time for the whole sky south of Declination $+30^\circ$ over a wider frequency range of 75--230\,MHz, allowing the calculation of the spectral index of such objects from MWA data alone, which will give a more uniform sampling of spatial scales than the range of instruments used in this paper. 

GRGs have very low brightness and will be fully resolved in any high resolution survey so it is possible that there is an undiscovered population of such sources; see e.g. \citet{2012ApJS..199...27S}. Assuming that GRGs are randomly distributed over the southern sky, and that GLEAM reaches the same sensitivity as the observations in this paper, we can only expect a few ($<10$) more objects of this brightness, so it will remain a rare population. Serendipitously, NGC\,1534 did not coincide with any bright (unrelated) sources at MWA frequencies; to best search for diffuse objects such as relic radio lobes, contaminating unrelated sources will need to be peeled. Assuming this can be done,
diffuse sources with peak surface brightness of $\approx80$\,mJy\,beam$^{-1}$ should be reliably detectable, which could unveil an even fainter population of relic sources.


\subsection{Relative lifetime and space density}

\citet{2010ApJ...713..398L} present analytic models 
for the time evolution of low-power radio galaxies at 1.4\,GHz, applied to a subset of 
local ($z=0.02$--0.3) Two-degree-Field Galaxy Redshift Survey \citep[2dFGRS;][]{1999RSPTA.357..105C} radio galaxies. We note that NGC\,1534 would not appear in this sample as it has a surface brightness below the 2dFGRS sample radio detection limit (see Figure~3 of \citealt{2010ApJ...713..398L}).

\begin{figure}
\begin{center}
\vbox{\includegraphics[width=8.5cm]{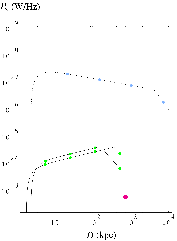}}
\caption{A reproduction of Figure~8 of \citet{2010ApJ...713..398L}, showing the evolutionary (1.4\,GHz-)power-size tracks for high-luminosity (FR-II) sources (dashed) and low-luminosity (FR-I) sources (solid). The dots (from left to right) on each curve indicate ages; for low-luminosity sources, of interest here, the green (lower) dots represent 1\,Myr, 10\,Myr, 10$^2$\,Myr, and 10$^3$\,Myr, for two slightly different models of radio galaxy evolution \citep[see][for more details]{2010ApJ...713..398L}. The (scaled 1.4-GHz) power and size of NGC\,1534 is overlaid with a large magenta circle.\label{fig:sadler_models}}
\end{center}
\end{figure}

In Figure~\ref{fig:sadler_models}, we reproduce Figure\,8 of \citet{2010ApJ...713..398L}, which shows the evolution of high-luminosity (FR-II) and low-luminosity (FR-I) sources in power and linear size according to the described models.
The power-size tracks are characterized by a slow increase in both size and luminosity over most of the source lifetime, followed by a rapid decline in luminosity at late times. Assuming that the extent of the lobes of NGC\,1534 is the same at 1.4\,GHz as observed in the MWA data, and that the CHIPASS upper limit is its true luminosity, we can overlay NGC\,1534 in the same plot. 
We see that it lies in the region of the $P$-$D$ diagram which corresponds to a low-power radio galaxy at the end of its life, with an age $>10^9$\,yr.
In this model the peak radio luminosity of the lobes of NGC\,1534 would have been at most a few times
10$^{24}$\,W\,Hz$^{-1}$, and perhaps a little lower. Note that the 1.4-GHz extent could be smaller, and the spectrum may curve downward at higher frequencies, so its position in Figure~\ref{fig:sadler_models} is likely to be an upper limit.

Since NGC\,1534 may have been caught in the rare dying phase, it is of interest to estimate the space density for such objects and hence the fraction of time an AGN might spend in this phase.
NGC\,1534 is the only double-lobed source of this angular size in the 4,000\,deg$^2$ area surveyed so far. Similar sources could have been detected at up to twice this distance, corresponding to a volume of $1.5\times10^6$\,Mpc$^3$. Thus the space density is $\approx7\times10^{-7}$\,Mpc$^{-3}$; our result is roughly seven times greater than that of \citet{2005AJ....130..896S} due to the higher surface brightness sensitivity of the MWA compared to SUMSS.

Using the bivariate luminosity function from \citet{2007MNRAS.375..931M}, we estimate that there are 15~AGN in this volume with radio luminosities greater than or equal to that of NGC\,1534 and identified with similar brightness elliptical galaxies.  Hence we can argue that normal AGN cannot spend more than 6\% of their lifetime in this phase if they all go through the same cycle. Of course it is possible that some AGN may not reach this linear size (e.g. those more confined in galaxy clusters) and may have a longer lifetime in the dying phase. To refine this estimate we would need to compare objects of similar linear size separating those with no evidence for ongoing energy input based on the absence of cores and hot spots and with a cut-off in their spectra. The spectral search will be possible with the GLEAM survey, now in progress. Determining the presence of cores and hotspots will require higher resolution surveys such as those to be carried out by the Australian SKA Pathfinder, and eventually SKA-mid.

\section{Conclusion}
\label{sec:conclusion}
We report the discovery of the lowest surface brightness radio galaxy yet detected, associated with the lenticular galaxy NCG\,1534. This represents only the seventh detection of large-scale radio lobes associated with a galaxy containing a dust lane. Additionally, NCG\,1534 has the steepest spectral index and lowest power of galaxies of this type. The lack of jets, hotspots and an obvious radio core associated with the optical host along with the non-detection of significant neutral hydrogen, despite the obvious dust lane, suggest that this is a true ``relic'' radio galaxy.

Deep observations at higher frequencies would be valuable in constraining the precise nature of the spectrum and in particular determine likely emission timescales and scenarios. Additionally, deep, high-resolution observations are required to better determine the core to lobe luminosity, as it is needed to confirm that the AGN is truly ``dead''. A measurement of the \textsc{Hi} mass would help determine whether there is any material available to feed the central black hole.

In the next few years LOFAR and the MWA will conduct a series of all-sky continuum radio surveys and will undertake a wide range of survey science projects \citep{norris13}. These surveys have the potential to uncover more systems of this type, although the expected source density remains low. With the advent of the SKA a decade hence, we will see a dramatic improvement in sensitivity to the population of ``relic'' radio galaxies and may perhaps detect them in sufficient numbers to start to undertake detailed statistical analysis of the population as a whole.
\section*{Acknowledgements}
We thank Drs Leith Godfrey and Lakshmi Saripalli for useful discussions on properties of dying radio galaxies. MJ-H is supported in this work through the Marsden Fund administered by the Royal Society of New Zealand.
This scientific work makes use of the Murchison Radio-astronomy Observatory, operated by CSIRO. We acknowledge the Wajarri Yamatji people as the traditional owners of the Observatory site.
Support for the MWA comes from the U.S. National Science Foundation (grants AST-0457585, PHY-0835713, CAREER-0847753, and AST-0908884), the Australian Research Council (LIEF grants LE0775621 and LE0882938), the U.S. Air Force Office of Scientic Research (grant FA9550-0510247), and the Centre for All-sky Astrophysics (an Australian Research Council Centre of Excellence funded by grant CE110001020). Support is also provided by the Smithsonian Astrophysical Observatory, the MIT School of Science, the Raman Research Institute, the Australian National University, and the Victoria University of Wellington (via grant MED-E1799 from the New Zealand Ministry of Economic Development and an IBM Shared University Research Grant).
The Australian Federal government provides additional support via the Commonwealth Scientific and Industrial Research Organisation (CSIRO), National Collaborative Research Infrastructure Strategy, Education Investment Fund, and the Australia India Strategic Research Fund, and Astronomy Australia Limited, under contract to Curtin University.
We acknowledge the iVEC Petabyte Data Store, the Initiative in Innovative Computing and the CUDA Center for Excellence sponsored by NVIDIA at Harvard University, and the International Centre for Radio Astronomy Research (ICRAR), a Joint Venture of Curtin University and The University of Western Australia, funded by the Western Australian State government.
This research has made use of the National Aeronautics and Space Administration (NASA) / Infrared Processing and Analysis Center (IPAC) Infrared Science Archive and the NASA/IPAC Extragalactic Database (NED) which are operated by the Jet Propulsion Laboratory, California Institute of Technology, under contract with NASA. This research has also made use of NASA's Astrophysics Data System.
The Digitized Sky Survey was produced at the Space Telescope Science Institute under US Government grant NAG W-2166 and is based on photographic data obtained using The UK Schmidt Telescope. The UK Schmidt Telescope was operated by the Royal Observatory Edinburgh, with funding from the UK Science and Engineering Research Council, until 1988 June, and thereafter by the Anglo-Australian Observatory.
Original plate material is copyright \copyright of the Royal Observatory Edinburgh and the Anglo-Australian Observatory.
The plates were processed into the present compressed digital form with their permission. 
SuperCOSMOS Sky Survey material is based on photographic data originating from the UK, Palomar and ESO Schmidt telescopes and is provided by the Wide-Field Astronomy Unit, Institute for Astronomy, University of Edinburgh.
\newcommand{\pasa}{PASA}
\newcommand{\aj}{AJ}
\newcommand{\apj}{ApJ}
\newcommand{\apjs}{ApJS}
\newcommand{\apjl}{ApJL}
\newcommand{\aap}{A{\&}A}
\newcommand{\aaps}{A{\&}AS}
\newcommand{\mnras}{MNRAS}
\newcommand{\araa}{ARAA}
\newcommand{\pasp}{PASP}
\newcommand{\nat}{Nature}
\bibliographystyle{mn2e_columbia}

\setlength{\labelwidth}{0pt}

\bibliography{refs}

\appendix

\bsp

\label{lastpage}

\end{document}